%% file: main.tex
\documentclass[runningheads]{llncs}
\usepackage{graphicx}
\usepackage{amsmath}
\usepackage{hhline}
\usepackage{booktabs}
\usepackage{color, colortbl}
\usepackage{tabularx}
\usepackage{hyperref}

\usepackage{multirow}
\definecolor{gray}{gray}{0.9}
\newcolumntype{a}{>{\columncolor{gray}}r}

\begin{document}
\title{Keypoints Localization for Joint Vertebra Detection and Fracture Severity Quantification}

\titlerunning{Keypoints for Vertebra Detection \& Fracture Quantification}

\authorrunning{M. Pisov, V. Kondratenko et al} 

\author{
    Maxim Pisov  \inst{1, 2} \and
    Vladimir Kondratenko \inst{1}  \and 
    Alexey Zakharov \inst{2,3}  \and 
    Alexey Petraikin \inst{4} \and
    Victor Gombolevskiy \inst{4} \and
    Sergey Morozov \inst{4} \and
    Mikhail Belyaev\inst{1}
}

\institute{
    Skolkovo Institute of Science and Technology, Moscow, Russia
    \and 
    Kharkevich Institute for Information Transmission Problems, Moscow, Russia
    \and
    Moscow Institute of Physics and Technology, Moscow, Russia
    \and 
    Research and Practical Clinical Center of Diagnostics and Telemedicine Technologies, Department of Health Care of Moscow, Russia
    \\
    \email{m.pisov@skoltech.ru}
}

\maketitle           
\begin{abstract}
Vertebral body compression fractures are reliable early signs of osteoporosis. Though these fractures are visible on Computed Tomography (CT) images, they are frequently missed by radiologists in clinical settings. Prior research on automatic methods of vertebral fracture classification proves its reliable quality; however, existing methods provide hard-to-interpret outputs and sometimes fail to process cases with severe abnormalities such as highly pathological vertebrae or scoliosis. 
We propose a new two-step algorithm to localize the vertebral column in 3D CT images and then to simultaneously detect individual vertebrae and quantify fractures in 2D. We train neural networks for both steps using a simple 6-keypoints based annotation scheme, which corresponds precisely to current medical recommendations. Our algorithm has no exclusion criteria, processes 3D CT in $2$ seconds on a single GPU and provides an intuitive and verifiable output. The method approaches expert-level performance and demonstrates state-of-the-art results in vertebrae 3D localization (the average error is $1$ mm), vertebrae 2D detection (precision is $0.99$, recall is $1$), and fracture identification (ROC AUC at the patient level is $0.93$). 

\end{abstract}

\setcounter{footnote}{0}

\input{content/introduction.tex}
\input{content/previous.tex}
\input{content/method.tex}
\input{content/data.tex}
\input{content/results.tex}
\input{content/discussion.tex}

\bibliographystyle{splncs04}
\bibliography{main}

\newpage
\section*{Supplementary materials}
\input{content/appendix.tex}

\end{document}

%% file: content/introduction.tex
\section{Introduction}    

Osteoporotic fractures are common in older adults and resulted in more than two million Disability Adjusted Life Years in Europe \cite{johnell2006estimate}. 
The presence of vertebrae fractures dramatically increases the probability of subsequent fractures \cite{klotzbuecher2000patients}; thus can be used as an early marker of osteoporosis.  Medical imaging, such as Computed Tomography (CT), is a useful tool to identify fractures \cite{lenchik2004diagnosis}. However, radiologists frequently miss fractures, especially if they are not specializing in musculoskeletal imaging; 
with the average error rate being higher than 50\% \cite{mitchell2017reporting}. 
At the same time, rapidly evolving low dose CT programs, e.g., for lung cancer, provide a solid basis for opportunistic screening of vertebral fractures. 

\begin{figure}[t]
    \begin{center}
      \includegraphics[width=\linewidth]{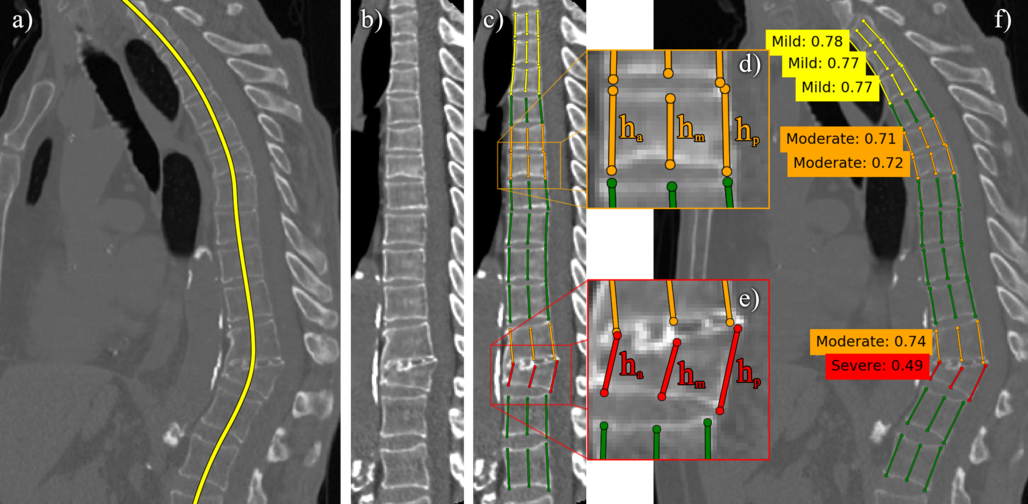}
      \caption{
      Overview of the proposed model.
      \textbf{Step 1}: a) localizing vertebrae centers in 3D CT (a sagittal projection is shown); 
      b) generating a new 2D image via spine `straightening'.
      \textbf{Step 2}: c) identifying key-points and the corresponding heights;
      d-e) a closer look at some vertebrae (colors denote the fracture severity).
      \textbf{Finally}: f) the original image with estimated fracture severities.
      }
      \label{fig:method}
    \end{center}
\end{figure} 

The medical image computing community thoroughly investigated fractures detection and/or classification on vertebrae-level \cite{roth2016deep,valentinitsch2019opportunistic,burns2017vertebral,antonio2018vertebra}, whole study-level \cite{tomita-rnn,zebra}, or jointly on both levels \cite{nicolaes-segmentation}, see Section \ref{sec:previous} for more details.
Many of these approaches require prior vertebrae detection \cite{antonio2018vertebra,valentinitsch2019opportunistic,nicolaes-segmentation}, or spine segmentation \cite{burns2017vertebral,roth2016deep,zebra}. Though both problems are active areas of research with prominent results, fractured vertebrae are the most complex cases for these algorithms \cite{sekuboyina2017attention}, and even good average detection/segmentation accuracy may not be sufficient for accurate fracture estimation. As a result, researchers had to exclude some studies from the subsequent fracture classification due to errors in prior segmentation \cite{valentinitsch2019opportunistic}, or due to scoliosis \cite{tomita-rnn}.

The second important issue is the mismatch between computer science problem statements and the radiological way to define fractures. The Genant scale \cite{genant} is a widely used medical criterion recommended by the International Osteoporosis Foundation \cite{iof_recs}. It relies on the measurements of  $h_a, h_m, h_p$ - the anterior, middle and posterior heights of vertebral bodies (Fig. \ref{fig:method}d, \ref{fig:method}e):
\begin{equation}
    \label{eq:genant}
    G = \frac{\min \{h_a, h_m, h_p\}}{\max \{h_a, h_m, h_p\}},
\end{equation}
$G$ values provide an easy to interpret continuous index, whereas existing methods are usually trained to predict a binary label extracted from radiological reports \cite{tomita-rnn,zebra} or multiclass labels based on threshold levels for $G$ \cite{valentinitsch2019opportunistic,burns2017vertebral}. A related problem is the interpretability of the methods' outputs. The only available information is the network's attention \cite{tomita-rnn} or a similar score \cite{nicolaes-segmentation} somehow related to the probability of fracture presence.  

\textbf{Our contribution} is two-fold. First, we propose a new method to identify the vertebral column in 3D CT and, as a consequence, reducing the problem to 2D by producing the corresponding mid-sagittal slice \cite{buckens2013intra} to measure $h_a, h_m, h_p$ for each vertebra (Fig. \ref{fig:method}a,b). Our method is trained to directly solve the localization problem rather than spine segmentation and demonstrates excellent localization quality with the average error less than $1$ mm. Also, it allows us to process all studies with no exceptions, including cases with severe scoliosis. 
Second, our method estimates six keypoints to detect each vertebra and estimate its heights $h_*$ simultaneously (Fig. \ref{fig:method}c-e), which results in excellent fracture classification quality with the area under ROC curve equal to $0.93$. The predictions are highly interpretable as they can be validated by a doctor using a simple ruler. 

%% file: content/previous.tex
\section{Previous work}
\label{sec:previous}
The automatic classification of vertebral fractures has received much attention from the medical image analysis community. A quantitative image analysis method was proposed in \cite{burns2017vertebral} to classify individual vertebra. First, the spinal column is segmented by an external method detecting intervertebral intervals. Then each vertebra is split into 17 sections to extract a set of simple features such as mean density from the segmentation mask. 
Finally, a support vector machine classifies vertebrae based on the obtained 51 features. The system provides excellent sensitivity (98.7\%) but quite low specificity (77.3\%).
A similar approach was used in \cite{valentinitsch2019opportunistic} where authors calculated  computer vision features such as histograms of oriented gradients from vertebra masks and achieved ROC AUC 0.88. A plain deep learning-based version of this two-step approach was proposed in \cite{antonio2018vertebra}, where classical ResNet was trained on 3-channel 2D images obtained from the prior segmentation mask by taking central sagittal, axial and coronal slices for each vertebra. 

It is important to note that all the methods above rely on prior segmentation, which may result in removing some cases with severe abnormalities. Indeed, the authors of \cite{valentinitsch2019opportunistic} reported that 11 cases out of 154 were excluded from the analysis due to incorrect prior spine segmentation largely caused by high-grade fractures.

This requirement was relaxed in several papers. In \cite{nicolaes-segmentation} the authors proposed a two-step pipeline for vertebrae detection: first, a segmentation neural network is used to generate pixel-level predictions (background, normal, fracture), then the predicted maps are aggregated. Instead of the whole spine mask, the authors used the ground-truth coordinates of vertebrae centroids to produce vertebrae-level predictions and achieved ROC AUC 0.93.  A simple idea was used in \cite{tomita-rnn}, where the authors selected the central sagittal slices as the spine is usually located in the middle of the image. In particular, they processed only 6.9 central slices per study (on average). As a result, this approach fails to identify fractures in patients with at least moderate scoliosis, and they had to exclude 156 out of 869 subjects from the analysis, primarily due to scoliosis. Though the average prevalence of scoliosis is 8.85\%, it positively correlated with age and increases from 10.95\% in 60-69 to 50\% in 90+ age groups \cite{kebaish2011scoliosis}, so this cohort can not be ignored in vertebral fractures screening. The classification method from \cite{tomita-rnn} consists of a ResNet34 which processes each of the central sagittal slices separately; then the obtained scores are aggregated by a simple LSTM network. 

Finally, an original approach was proposed in \cite{zebra}. Though the method also relies on external spine segmentation, the mask is used to extract the spinal cord and create a new virtual sagittal slice. Next, small patches are extracted from this slice and classified by a convolutional network; finally, a recurrent neural network (RNN) is used to aggregate the predictions from each patch. Although the training database is the largest among the reviewed works (consisting of 1673 cases), the model achieves 83.9\% sensitivity (with 93.8\% specificity), likely due to poor study-level binary annotation extracted from the radiological reports.

%% file: content/method.tex
\section{Method}
\label{sec:method}

The majority of existing methods are two-step pipelines: first, the spine is localized or segmented; second, individual vertebrae are processed to identify fractures. We follow the same scheme with two major goals:
\begin{enumerate}
	\item Replace the first part by a more task-specific alternative to avoid the exclusion of any case from the analysis due to segmentation failures.
 	\item Create a method capable of directly working with Genant fracture reporting
\end{enumerate}

The second, more important goal dictated the annotation protocol, which affects both steps of our method. Following \cite{buckens2013intra}, raters were instructed to find the mid-sagittal slice for each of the visible vertebrae and annotate six keypoints to measure anterior, middle and posterior heights (Fig. 1d, 1e).

In the first part of our pipeline we predict 3D coordinates of the middle height keypoints using a 3D fully convolutional neural network with soft-argmax activation \cite{soft-argmax}. The predicted coordinates are then used to localize the spine and to select a sagittal plane which contains all the vertebrae - an idea introduced in \cite{bromiley2016fully} and later applied in \cite{zebra}.
However, we cannot simply choose a sagittal plane on the original image, because such a plane might simply not exist, e.g. for patients with severe spinal scoliosis.
For this reason the obtained 3D curve is used to `straighten' the spine and generate a new 2D slice (Sec. \ref{sec:straightening}). 
It is worth noting that though the input is a 3D CT scan, the annotation is intrinsically bidimensional, so this dimensionality reduction is essential. The key advantage of our approach is that we directly use final annotation to find the most appropriate 2D representation of the original 3D image.  

The second part of our network processes the obtained 2D image in order to localize each vertebra and predict positions of six keypoints. Given the fact that the number of vertebrae for a given input image is not known a priori, a natural solution is to use object detection techniques in order to make vertebra-level predictions without the need for additional postprocessing. Our major insight is that 
directly predicting keypoints coordinates relative to the bounding box center shows a dramatic increase in model quality (Sec. \ref{sec:detection}). We also propose a combined loss function to enforce good quality of $G$ index in addition to localization and heights estimation parts. 

Finally the predictions of the second network are mapped back to 3D in order to calculate the heights $h_a, h_m, h_p$ and the $G$ index. 

\subsection{Spine straightening}
\label{sec:straightening}
We use a 3D UNet-like \cite{resunet} architecture to predict a 2D probability map on each axial slice, followed by the 2D soft-argmax operation \cite{soft-argmax} to obtain spatial coordinates of the vertebral body central line. We train our network by optimizing \textit{mean absolute error} between the predicted points and the ones smoothly interpolated from the annotation of middle height endpoints (Fig. \ref{fig:target}a). 

\begin{figure}[t]
    \begin{center}
      \includegraphics[width=1.\linewidth]{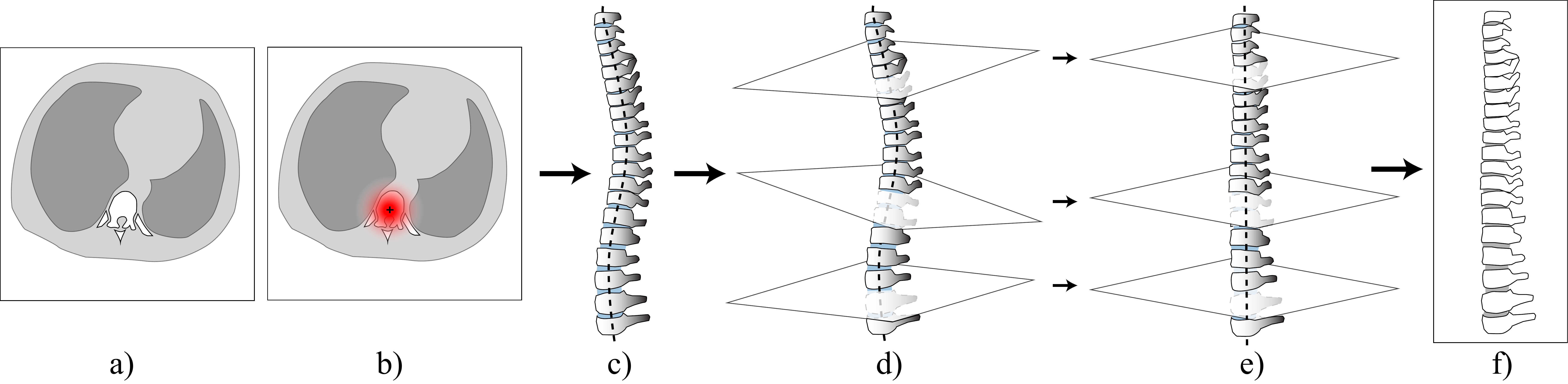}
      \caption{
      The spine straightening pipeline: 
      a) a single axial slice; 
      b) an axial slice with the probability heatmap (red), the cross indicates the resulting point after the soft-argmax operation; 
      c) the combined points from each slice result in a 3D curve; 
      d) planes, orthogonal to the curve (for better visualization most planes are omitted);
      e) a straightened vertebral column (the planes become parallel);
      f) the new central sagittal plane.
      }
      \label{fig:straighten}
    \end{center}
\end{figure}

By combining the predictions for each axial plane, we obtain a 3D curve.
We then interpolate the image onto a new 3D grid on which the obtained curve becomes a straight vertical line. The grid is constructed in such a way, so that the planes normal to the curve become parallel (Fig. \ref{fig:straighten}d-e). Finally we select a new sagittal plane where all vertebrae are visible. Fig. \ref{fig:straighten} shows a detailed illustration.

\subsection{Vertebrae-level predictions}
\label{sec:detection}
 
The object detection step of our method is mainly based on YOLO9000 \cite{yolo9000} and YOLOv3 \cite{yolo-v3} with several modifications, in order to adapt it to the specifics of the given task. 
Similarly to YOLO9000, our architecture consists of a single Region Proposal Network (RPN), that makes predictions relative to a set of anchor boxes, and doesn't require ROI-pooling \cite{faster-rcnn} and any further refinement.
Additionally, YOLOv3 uses Feature Pyramid Networks (FPNs) \cite{fpn}, which enables it to make accurate predictions on various scales. As we know that the range of shapes a vertebra can have is quite narrow, we use a simple UNet-based \cite{unet} architecture instead. Finally, as opposed to the YOLO family, our RPN makes predictions in the original resolution, because we favor accuracy over speed.

\begin{figure}
    \begin{center}
      \includegraphics[width=.85\linewidth]{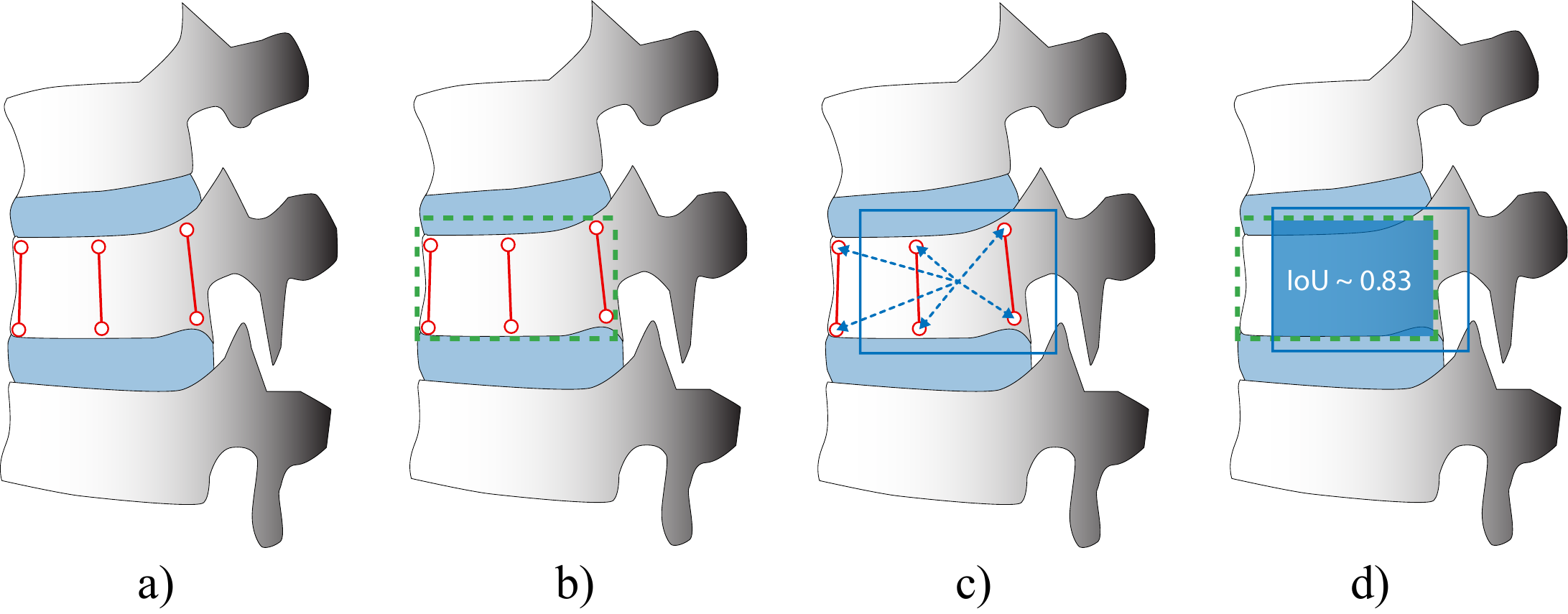}
      \caption{Target generation steps: a) example of an annotated vertebra; b) the generated axis-aligned bounding box (green, dashed); c) the \textbf{keypoints' coordinates} relative to anchor box' center; d) the \textbf{objectness} $O$ is 1, if IoU between the boxes is greater than 0.5}
      \label{fig:target}
    \end{center}
\end{figure}

The \textbf{target generation} pipeline is shown in Fig. \ref{fig:target}. Note, that, because our goal is to assign 6 keypoints per vertebra, we don't really need to predict bounding boxes. Instead, we generate \textit{axis aligned bounding boxes} (Fig. \ref{fig:target}b,d) to calculate \textit{intersection-over-union} (IoU) at both train and test time.
We encode the target keypoints coordinates by using the same scale- and shift-invariant encoding as in \cite{faster-rcnn}:
\begin{equation}
    \label{eq:keypoints}
    e^x = (g^x - a^x) / a^w; \quad e^y = (g^y - a^y) / a^h,   
\end{equation}
where $(g^x, g^y), (e^x, e^y)$ are the global and encoded coordinates of a given keypoint respectively, $(a^x, a^y)$ - is the center of an anchor box, and $(a^w, a^h)$ are its width and height (Fig. \ref{fig:target}). We selected the following anchor boxes' scales: $17,\, 23,\, 28,\, 35\; mm$ and aspect ratios: $0.8, \, 1.1, \, 1.3, \, 2$.

Finally, we propose the following \textbf{loss function} to train our second network:
\begin{equation}
    \label{eq:loss}
    L = BCE(\hat o, o) + 
     \frac{1}{\sum\limits_i I[o_i = 1]}\sum\limits_i
      \frac{I[o_i = 1]}{G_i} \cdot MAE(\hat e_i, e_i),
\end{equation}
where 
both sums are calculated over all vertebrae in the training batch,
$BCE$ is the standard \textit{log-loss} between real ($o$) and predicted ($\hat o$) objectness (Fig. \ref{fig:target}d),
$MAE$ is the \textit{mean absolute error} between real ($e_i$) and predicted ($\hat e_i$) encoded keypoints' coordinates (\ref{eq:keypoints}) for the i-th vertebra and
$G_i$ is the respective Genant score (\ref{eq:genant}).

%% file: content/data.tex
\section{Data}

Our dataset consists of 100 chest CT. It represents a randomly selected subset of a publicly available dataset\footnote{\url{https://mosmed.ai/datasets/ct\_lungcancer\_500}}.
The images have various voxel spacing ranging from $.5 \times .5 \times .8$mm to $1 \times 1 \times .8$mm and different numbers of visible vertebrae: from 10 to 15.

The data was annotated by 7 experts with $1$ to $5$ years of experience in radiology and a board-certified radiologist with 12 years of experience in the field. In total the dataset contains 1268 annotated vertebrae with 2-3 annotations per vertebra. The distribution of vertebral fractures is the following: 125 mild, 80 moderate, 17 severe deformations and 1046 normal vertebrae.
Patient-wise we have a somewhat balanced distribution with 30, 16, 41 and 13 patients with none, mild, moderate and severe deformations respectively. 

%% file: content/results.tex
\section{Results}
\subsection{Experimental setup}

In all of our experiments the only preprocessing we use is intensity normalization to zero mean and unit variance. 

We trained our \textbf{spine straightening} network with the Adam \cite{adam} optimizer with default parameters ($\beta_1 = 0.9, \beta_2 = 0.999$) and a learning rate of $10^{-3}$, which showed the fastest convergence rate. As the architecture operates on whole 3D images, we reduce the images' resolution to a spacing of $3 \times 3 \times 3$ mm, the predicted curve is then linearly interpolated to the original resolution. In such a setting the network reached convergence after approx. 10k batches of size 3.

Similarly we trained the \textbf{vertebrae detection} network for 4k iterations with batches of size 30.
We start with a learning rate of $10^{-4}$ at the early stages of training and decrease it by a factor of 2 after 1k, 1.4k and 2k iterations, because gradually decreasing the learning rate enabled the model to reach better optima.

We reported results obtained using 5-fold cross-validation. As we have multiple annotations per study, we also report the inter-expert variability. To obtain patient-level predictions, we use the most severe fracture among all the vertebrae, which is equivalent to taking the minimal Genant score.

\subsection{Method performance}
We report the localization quality of the first step of our method in Tab. \ref{tab:localization_detection}, left. To calculate these numbers, we found the closest annotated vertebrae in 3D. Also, we report 2D detection metrics for the second network, see Tab. \ref{tab:localization_detection}, right. 

\begin{table}
\caption{
    \label{tab:localization_detection}
    Vertebral body centers localization
    and vertebrae detection metrics. Columns with white background denote the average (std) number for all vertebrae, the ones with grey background - for \textit{Moderately} and \textit{Severely} fractured vertebrae ($G\le 0.74$). The ground truth $G$-index is not defined for false positives, so Precision is reported for all vertebrae only.
}
\begin{center}
\begin{tabular}{crarac}
    \toprule 
   & \multicolumn{2}{c}{Localization, mm} &     \multicolumn{2}{c}{Recall} &
    Precision  \\ 
    \cmidrule(lr){2-3} \cmidrule(lr){4-5} \cmidrule(lr){6-6}   
    Proposed & 
        \enskip 0.97	(0.64)  &
        \enskip 1.14	(0.69) &
        \enskip 0.997 (0.003) &
        \enskip 1.000 (0.000)  &
        \enskip 0.993 (0.002) \\
        
    Experts & 1.01	(0.98)  & 1.17	(0.78)
             &0.983 (0.006)& 0.973 (0.013)
             & \enskip 0.996 (0.001) \\
    \bottomrule
\end{tabular}
\end{center}
\end{table}

To analyze the performance of vertebrae fractures classification, we report metrics for two threshold values of $G$ following the radiological definition of severity \cite{genant}, see Tab. \ref{tab:vertebra}. We assume that the most relevant problem for chest CT is the identification of at least Moderate fractures ($G\le0.74$) as healthy vertebrae in the thoracic spine are wedged, so normal variation can be misclassified as a Mild fracture ($0.74<G\le0.8$) \cite{lenchik2004diagnosis}. 
The obtained results are close to human-level performance on our dataset and comparable with other works. Similar values of ROC AUC were obtained at vertebra ($0.88$ \cite{valentinitsch2019opportunistic}, $0.93$ \cite{nicolaes-segmentation}) and patient levels ($0.92$ \cite{tomita-rnn}). 

However, we cannot directly compare the performance with other works due to several factors. First, different definitions of fractured vertebrae are used across papers. Second, abdominal and pelvis CTs differ from chest CT in the number of visible vertebrae and anatomy (e.g., the above-mentioned fact concerning a higher error rate in chest CT). This fact motivates us to release our annotation and provide the community test data for further development\footnote{\url{https://github.com/neuro-ml/vertebral-fractures-severity}}.

Fig. \ref{fig:method} shows an example of the inference process on an image from the dataset. Due to limited space, we refer the interested reader to the supplementary materials for a broader set of examples. The overall inference takes under $2$ seconds on Nvidia GTX 980ti, with an approximately equal time required for spine localization and all the subsequent steps (including spine straightening).

\begin{table}[t]
\caption{
    \label{tab:vertebra}
    Binary classification metrics for various grades of fractures: at least \textit{Mild} ($G \le 0.8$) and at least \textit{Moderate} ($G \le 0.74)$.
    All numbers are given as mean (std).
    Columns with white background denote vertebrae-level predictions, the ones with grey background - patient-level predictions.
}
\begin{center}
\begin{tabular}{ccrarara}
    \toprule 
    Grade & $G$ by & \multicolumn{2}{c}{ROC AUC} & \multicolumn{2}{c}{Specificity} &
    \multicolumn{2}{c}{Sensitivity} \\ 
    \cmidrule(lr){1-1} \cmidrule(lr){2-2} \cmidrule(lr){3-4} \cmidrule(lr){5-6} \cmidrule(lr){7-8}

   \multirow{2}{*}{Mild} &
   Proposed
   &\enskip
        .87 (.02) & \enskip
        .93 (.03)   & \enskip
        .93 (.01)   & \enskip
        .68 (.08)   & \enskip
        .65 (.03)   & \enskip
        .94 (.03)  \\
    
    &Experts \quad &
        .91 (.02) &
        .91 (.05) & 
        .91 (.01) &
        .60 (.11) & 
        .70 (.05) &
        .90 (.05) \\ \hline 

    \multirow{2}{*}{Moderate\;} &
    Proposed \quad &
        .94 (.02) &
        .93 (.03) &
        .98 (.01) &
        .86 (.05) &
        .75 (.05) &
        .84 (.05) \\
        
    &Experts \quad &\
        .98 (.01) &
        .95 (.03) & 
        .97 (.01) &
        .83 (.07) &
        .79 (.06) & 
        .88 (.07) \\

    \bottomrule
\end{tabular}
\end{center}
\end{table}

%% file: content/discussion.tex
\section{Discussion}

We proposed a new method for automatic identification of vertebrae-level fractures classification using the Genant score, which approaches, and in some cases surpasses, the inter-expert variability. Our analysis of examples on which the model performs poorly (some of which can be found in the appendix) shows that the experts' variability in these cases is also unusually high.
Note that our method can be easily adapted to more common 2D X-Ray images by simply dropping the first network. Also, similarly to Mask R-CNN \cite{mask-rcn}, we can extend our method to predict additional metadata for each vertebra, such as labels or segmentation masks.  Particularly, it would be interesting to validate our method on the challenging localization and labeling dataset \cite{glocker2013vertebrae}.

%% file: content/appendix.tex
\subsection*{Method performance examples}

\begin{figure}
    \begin{center}
      \includegraphics[width=\linewidth]{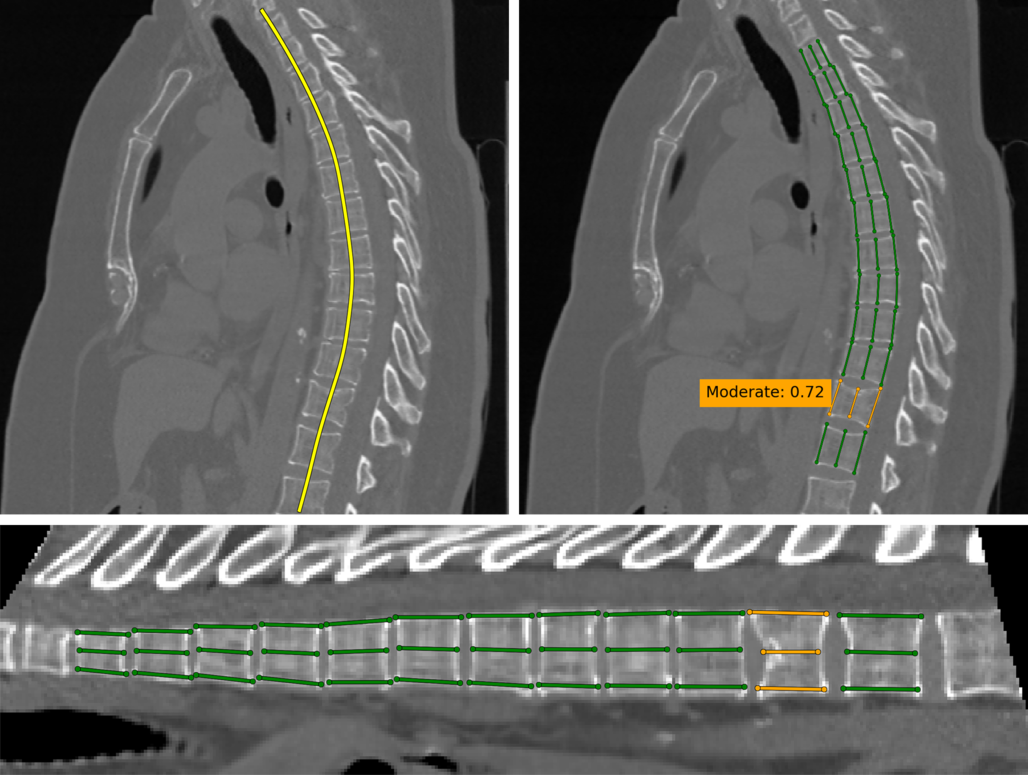}
    \end{center}
    \caption{A simple example from the dataset. Colors denote the fractures severity.}
\end{figure}

\begin{figure}
    \begin{center}
      \includegraphics[width=\linewidth]{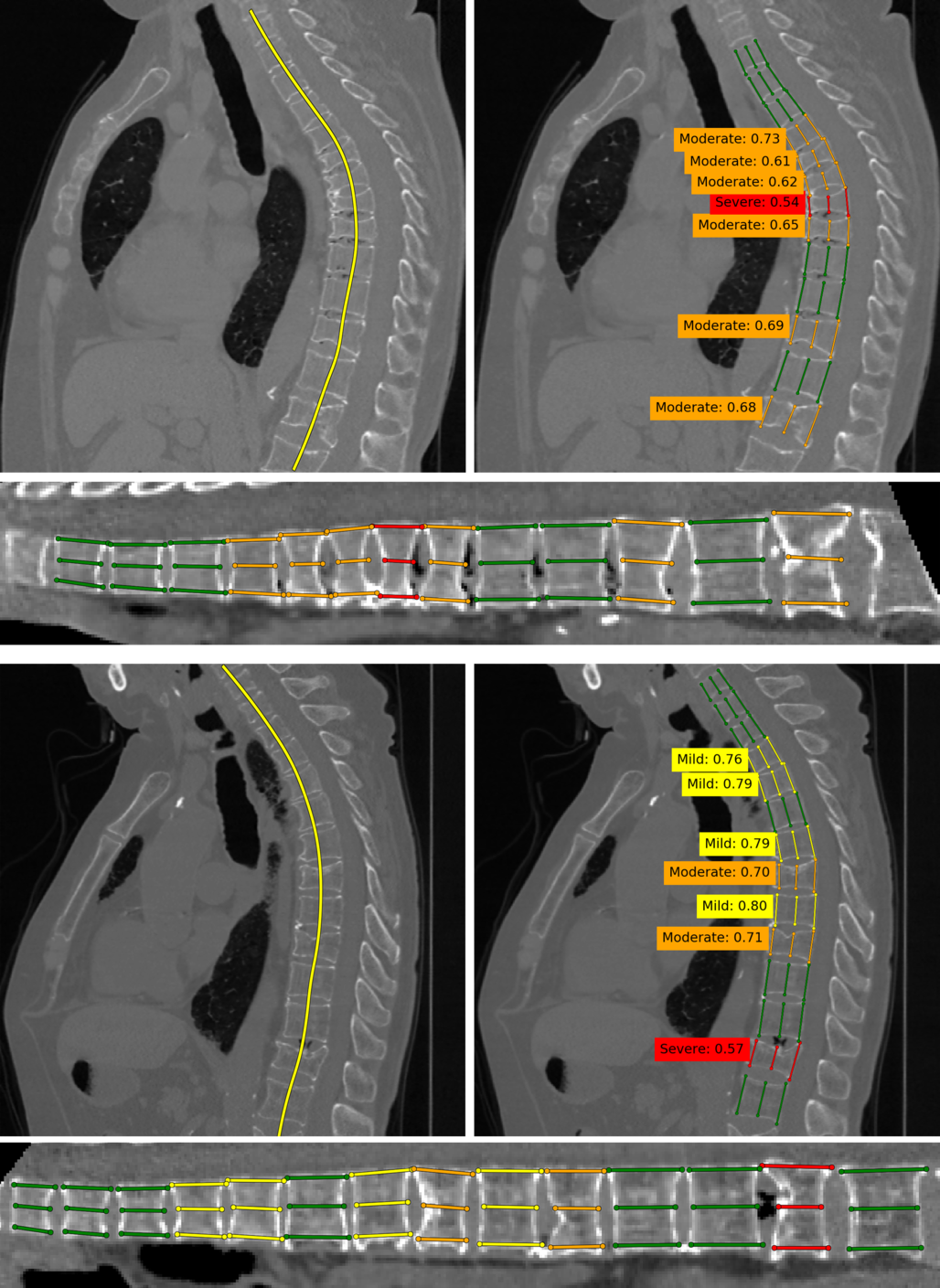}
    \end{center}
    \caption{Two hard examples from the dataset. Colors denote the fractures severity.}
\end{figure}

\begin{figure}
    \begin{center}
      \includegraphics[width=\linewidth]{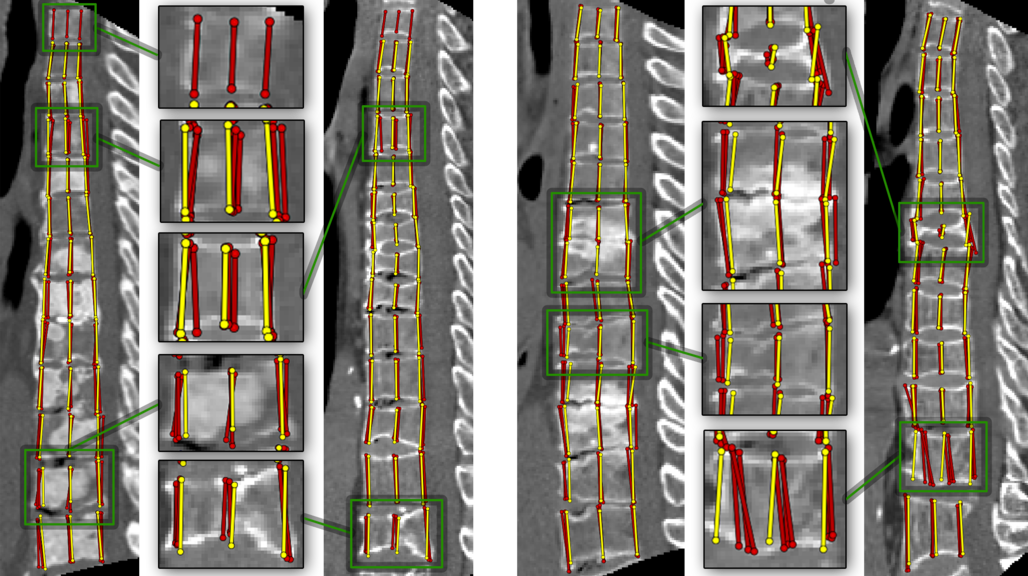}
    \end{center}
    \caption{Examples on which our model (yellow) performs poorly. Note the variability of the expert annotation (red).}
\end{figure}

%% file: main.bbl
\begin{thebibliography}{10}
\providecommand{\url}[1]{\texttt{#1}}
\providecommand{\urlprefix}{URL }
\providecommand{\doi}[1]{https://doi.org/#1}

\bibitem{antonio2018vertebra}
Antonio, C.B., Bautista, L.G.C., Labao, A.B., Naval, P.C.: Vertebra fracture
  classification from 3d ct lumbar spine segmentation masks using a
  convolutional neural network. In: Asian Conference on Intelligent Information
  and Database Systems. pp. 449--458. Springer (2018)

\bibitem{zebra}
Bar, A., Wolf, L., Amitai, O.B., Toledano, E., Elnekave, E.: Compression
  fractures detection on ct. In: Medical Imaging 2017: Computer-Aided
  Diagnosis. vol. 10134, p. 1013440. International Society for Optics and
  Photonics (2017)

\bibitem{bromiley2016fully}
Bromiley, P.A., Kariki, E.P., Adams, J.E., Cootes, T.F.: Fully automatic
  localisation of vertebrae in ct images using random forest regression voting.
  In: International Workshop on Computational Methods and Clinical Applications
  for Spine Imaging. pp. 51--63. Springer (2016)

\bibitem{buckens2013intra}
Buckens, C.F., de~Jong, P.A., Mol, C., Bakker, E., Stallman, H.P., Mali, W.P.,
  van~der Graaf, Y., Verkooijen, H.M.: Intra and interobserver reliability and
  agreement of semiquantitative vertebral fracture assessment on chest computed
  tomography. PloS one  \textbf{8}(8) (2013)

\bibitem{burns2017vertebral}
Burns, J.E., Yao, J., Summers, R.M.: Vertebral body compression fractures and
  bone density: automated detection and classification on ct images. Radiology
  \textbf{284}(3),  788--797 (2017)

\bibitem{iof_recs}
Genant, H.K., Bouxsein, M.L.: {Vertebral Fracture Initiative}: Executive
  summary.
  \url{https://www.iofbonehealth.org/sites/default/files/PDFs/IOF\_VFI-Executive\_Summary-English.pdf}
  (2011)

\bibitem{genant}
Genant, H.K., Wu, C.Y., Van~Kuijk, C., Nevitt, M.C.: Vertebral fracture
  assessment using a semiquantitative technique. Journal of bone and mineral
  research  \textbf{8}(9),  1137--1148 (1993)

\bibitem{glocker2013vertebrae}
Glocker, B., Zikic, D., Konukoglu, E., Haynor, D.R., Criminisi, A.: Vertebrae
  localization in pathological spine ct via dense classification from sparse
  annotations. In: International conference on medical image computing and
  computer-assisted intervention. pp. 262--270. Springer (2013)

\bibitem{mask-rcn}
He, K., Gkioxari, G., Doll{\'a}r, P., Girshick, R.: Mask r-cnn. In: Proceedings
  of the IEEE international conference on computer vision. pp. 2961--2969
  (2017)

\bibitem{johnell2006estimate}
Johnell, O., Kanis, J.: An estimate of the worldwide prevalence and disability
  associated with osteoporotic fractures. Osteoporosis international
  \textbf{17}(12),  1726--1733 (2006)

\bibitem{kebaish2011scoliosis}
Kebaish, K.M., Neubauer, P.R., Voros, G.D., Khoshnevisan, M.A., Skolasky, R.L.:
  Scoliosis in adults aged forty years and older: prevalence and relationship
  to age, race, and gender. Spine  \textbf{36}(9),  731--736 (2011)

\bibitem{adam}
Kingma, D.P., Ba, J.: Adam: A method for stochastic optimization. arXiv
  preprint arXiv:1412.6980  (2014)

\bibitem{klotzbuecher2000patients}
Klotzbuecher, C.M., Ross, P.D., Landsman, P.B., Abbott~III, T.A., Berger, M.:
  Patients with prior fractures have an increased risk of future fractures: a
  summary of the literature and statistical synthesis. Journal of bone and
  mineral research  \textbf{15}(4),  721--739 (2000)

\bibitem{lenchik2004diagnosis}
Lenchik, L., Rogers, L.F., Delmas, P.D., Genant, H.K.: Diagnosis of
  osteoporotic vertebral fractures: importance of recognition and description
  by radiologists. American Journal of Roentgenology  \textbf{183}(4),
  949--958 (2004)

\bibitem{fpn}
Lin, T.Y., Doll{\'a}r, P., Girshick, R., He, K., Hariharan, B., Belongie, S.:
  Feature pyramid networks for object detection. In: Proceedings of the IEEE
  conference on computer vision and pattern recognition. pp. 2117--2125 (2017)

\bibitem{soft-argmax}
Luvizon, D.C., Tabia, H., Picard, D.: Human pose regression by combining
  indirect part detection and contextual information. Computers \& Graphics
  \textbf{85},  15--22 (2019)

\bibitem{resunet}
Milletari, F., Navab, N., Ahmadi, S.A.: V-net: Fully convolutional neural
  networks for volumetric medical image segmentation. In: 3D Vision (3DV), 2016
  Fourth International Conference on. pp. 565--571. IEEE (2016)

\bibitem{mitchell2017reporting}
Mitchell, R., Jewell, P., Javaid, M., McKean, D., Ostlere, S.: Reporting of
  vertebral fragility fractures: can radiologists help reduce the number of hip
  fractures? Archives of osteoporosis  \textbf{12}(1), ~71 (2017)

\bibitem{nicolaes-segmentation}
Nicolaes, J., Raeymaeckers, S., Robben, D., Wilms, G., Vandermeulen, D.,
  Libanati, C., Debois, M.: Detection of vertebral fractures in ct using 3d
  convolutional neural networks. arXiv preprint arXiv:1911.01816  (2019)

\bibitem{yolo9000}
Redmon, J., Farhadi, A.: Yolo9000: better, faster, stronger. In: Proceedings of
  the IEEE conference on computer vision and pattern recognition. pp.
  7263--7271 (2017)

\bibitem{yolo-v3}
Redmon, J., Farhadi, A.: Yolov3: An incremental improvement. CoRR
  \textbf{abs/1804.02767} (2018), \url{http://arxiv.org/abs/1804.02767}

\bibitem{faster-rcnn}
Ren, S., He, K., Girshick, R., Sun, J.: Faster r-cnn: Towards real-time object
  detection with region proposal networks. In: Advances in neural information
  processing systems. pp. 91--99 (2015)

\bibitem{unet}
Ronneberger, O., Fischer, P., Brox, T.: U-net: Convolutional networks for
  biomedical image segmentation. In: International Conference on Medical image
  computing and computer-assisted intervention. pp. 234--241. Springer (2015)

\bibitem{roth2016deep}
Roth, H.R., Wang, Y., Yao, J., Lu, L., Burns, J.E., Summers, R.M.: Deep
  convolutional networks for automated detection of posterior-element fractures
  on spine ct. In: Medical Imaging 2016: Computer-Aided Diagnosis. vol.~9785,
  p. 97850P. International Society for Optics and Photonics (2016)

\bibitem{sekuboyina2017attention}
Sekuboyina, A., Kuka{\v{c}}ka, J., Kirschke, J.S., Menze, B.H., Valentinitsch,
  A.: Attention-driven deep learning for pathological spine segmentation. In:
  International Workshop and Challenge on Computational Methods and Clinical
  Applications in Musculoskeletal Imaging. pp. 108--119. Springer (2017)

\bibitem{tomita-rnn}
Tomita, N., Cheung, Y.Y., Hassanpour, S.: Deep neural networks for automatic
  detection of osteoporotic vertebral fractures on ct scans. Computers in
  biology and medicine  \textbf{98},  8--15 (2018)

\bibitem{valentinitsch2019opportunistic}
Valentinitsch, A., Trebeschi, S., Kaesmacher, J., Lorenz, C., L{\"o}ffler, M.,
  Zimmer, C., Baum, T., Kirschke, J.: Opportunistic osteoporosis screening in
  multi-detector ct images via local classification of textures. Osteoporosis
  international  \textbf{30}(6),  1275--1285 (2019)

\end{thebibliography}
